\documentclass[preprint,aps]{revtex4}
\usepackage{mathrsfs}

\usepackage{graphicx}% Include figure files
\usepackage{dcolumn}% Align table columns on decimal point
\usepackage{bm}% bold math
\usepackage{amsmath}
\usepackage{graphicx}

\begin{document}

\title{Orbital-Selective Spin Texture and its Manipulation in a Topological Insulator}

\author{Zhuojin Xie$^{1,\dag}$, Shaolong He$^{1,\dag}$, Chaoyu Chen$^{1}$, Ya Feng$^{1}$, Hemian Yi$^{1}$, Aiji Liang$^{1}$, Lin Zhao$^{1}$, Daixiang Mou$^{1}$, Junfeng He$^{1}$, Yingying Peng$^{1}$, Xu Liu$^{1}$, Yan Liu$^{1}$, Guodong Liu$^{1}$, Xiaoli Dong$^{1}$, Li Yu$^{1}$, Jun Zhang$^{1}$, Shenjin Zhang$^{3}$, Zhimin Wang$^{3}$, Fengfeng Zhang$^{3}$, Feng Yang$^{3}$, Qinjun Peng$^{3}$, Xiaoyang Wang$^{3}$, Chuangtian Chen$^{3}$, Zuyan Xu$^{3}$ and X. J. Zhou$^{1,2}$}

\affiliation{
\\$^{1}$Beijing National Laboratory for Condensed Matter Physics, Institute of Physics, Chinese Academy of Sciences, Beijing 100190, China
\\$^{2}$Collaborative Innovation Center of Quantum Matter, Beijing, China
\\$^{3}$Technical Institute of Physics and Chemistry, Chinese Academy of Sciences, Beijing 100190, China.
\\$^{\dag}$These people contribute equally to the present work.
%%\\$^{*}$Corresponding author: XJZhou@aphy.iphy.ac.cn.
}

\date{May 20, 2013}

%%\begin{abstract}

%\pacs{}% PACS, the Physics and Astronomy

\maketitle

\newpage

%%%\section{INTRODUCTION}

{\bf
Topological insulators represent a new quantum state of matter that are insulating in the bulk but metallic on the edge or surface. In the Dirac surface state, it is well-established that the electron spin is locked with the crystal momentum. Here we report a new phenomenon of the spin texture locking with the orbital texture in a topological insulator Bi$_2$Se$_3$.  We observe light-polarization-dependent spin texture of both the upper and lower Dirac cones that constitutes strong evidence of the orbital-dependent spin texture in Bi$_{2}$Se$_{3}$. The different spin texture detected in variable polarization geometry is the manifestation of the spin-orbital texture in the initial state combined with the photoemission matrix element effects. Our observations provide a new orbital degree of freedom and a new way of light manipulation in controlling the spin structure of the topological insulators that are important for their future applications in spin-related technologies.
}

\newpage

\noindent{\bf\large Introduction}

\noindent Topological insulators represent a new quantum state of matter which are insulating in their bulk interior but metallic on the topologically protected edge or surface states\cite{Fu1,Qi1,Hasan1}. In the conducting surface state of the three-dimensional topological insulators, the electron spin is locked to its crystal momentum, forming a unique helical spin texture\cite{Qi1}. Such a unique spin texture can suppress electron backscattering because of the protection of the time reversal symmetry, thus making the surface state robust against external perturbations\cite{Haijun1,Chaoyu}. Topological insulators are therefore promising for applications in spin-related electronics. The nontrivial spin texture of topological insulators has been intensively explored theoretically\cite{Zhangwei,Fu2,Basak,Louie}, and the spin-momentum locking has been demonstrated by a number of experiments\cite{Hsieh,Souma,Xu,Pan,Jozwiak,miyamoto}.  On the other hand, caution should be taken in the interpretation of spin-resolved photoemission results on topological insulators because of the strong spin-orbit coupling and the spin-dependent transition matrix element effects involved in the photoemission process. It has been shown that the detected spin texture of photoelectrons can be completely different from that of the initial states\cite{Park,Lanzara}. It was suggested that the usually accepted spin-conservation scenario for the spin-resolved photoemission may become invalid in these materials\cite{Lanzara}. Meanwhile, it was proposed very recently that the spin texture may be coupled with the orbital texture in topological insulators, which can only be examined directly by spin- and angle-resolved photoemission spectroscopy (SARPES) combined with polarization-variable light source\cite{Haijun2}. Such a new spin-orbital locking picture, if proved,  can provide new insight on the spin texture formation and paves a way for light manipulation on the orbital-selective spin textures in topological insulators.

Angle-resolved photoemission spectroscopy (ARPES) is a powerful tool to study the electronic structure of materials\cite{Shen1} and it has played a key role in discovering three-dimensional topological insulators\cite{Hasan1st,YLChenScience}. ARPES can measure not only the energy and momentum of electrons, but also the orbital characters of the measured electronic state\cite{Shen1}. Utilizing different linearly-polarized light, different orbital textures of the Bi$_{2}$Se$_{3}$ Dirac surface state can be probed\cite{caoyue1}. The spin-resolved ARPES goes one step further to measure not only the electron energy and momentum but also its spin state, providing complete information to describe the electronic states in a solid\cite{Dil1,Dil2,Shen1}. It has been employed to directly reveal the nontrivial spin texture of the topological insulators\cite{Hsieh,Souma,Xu,Pan,Jozwiak,miyamoto}. The spin-resolved ARPES, combined with the variable light polarization, has been proposed to be a necessary and most suitable experimental tool to detect the spin-orbital texture predicted in the Bi$_{2}$Se$_{3}$ surface state\cite{Haijun2}. In this report, we present strong evidence of the spin-orbital texture in Bi$_{2}$Se$_{3}$ by using our newly-developed SARPES system with polarization-variable vacuum ultraviolet (VUV) laser source. We demonstrate that, by switching the incident light from {\it p} to {\it s} polarization geometry, the spin texture for the upper Dirac cone changes from the left-handed to the right-handed chirality while it keeps the same right-handed chirality for the lower Dirac cone. In particular, in the {\it s} polarization geometry, our SARPES measurements reveal a surprising helical spin texture that the upper and lower Dirac cones share the same right-handed chirality.  This is distinct from the usual spin texture in the topological insulators where the upper and the lower Dirac cones show opposite spin chirality.  Our results provide strong experimental evidence for the orbital-selective spin texture in the Bi$_{2}$Se$_{3}$ topological insulator\cite{Haijun2}.\\

\noindent{\bf\large Results}

\noindent{\bf Fermi surface of Bi$_{2}$Se$_{3}$.} By utilizing polarized light source and taking advantage of the photoemission matrix element effect, ARPES can selectively excite and probe orbitals involved in the measured electronic states\cite{Shen1}. Figure 1 shows the ARPES spectral weight distribution of the Bi$_{2}$Se$_{3}$ topological insulator at different binding energies measured by {\it s} (upper panels) and {\it p} polarization (lower panels) geometries (for measurement geometries, see Methods below). In the {\it s} polarization measurements, the spectral weight above the Dirac point is strong along the k$_\text{x}$ direction but suppressed along the k$_\text{y}$ direction (Fig. 1a and 1b). Such an intensity distribution is reversed for the lower Dirac cone where the intensity along the k$_y$ is strong but suppressed along the k$_\text{x}$ direction (Fig. 1d). In the {\it p} polarization measurements, the spectral intensity is more uniform with slight variation in the measured constant energy contours (Fig. 1e and 1f). These observations are consistent with the previous report\cite{caoyue1} that can be well understood in terms of different components of the {\it p} orbitals probed under different polarization geometries due to the photoemission matrix element effect (Fig. 2). These results indicate that, in the {\it p} polarization geometry, electrons with the out-of-plane {\it p}$_\text{z}$ orbital characters are mainly selected and excited both for the upper and lower Dirac cones (Fig. 2b). In the {\it s} polarization geometry, on the other hand, electrons with dominantly in-plane tangential orbital texture are excited in the upper Dirac cone, while electrons with dominantly radial orbital texture are excited in the lower Dirac cone, as shown in Fig. 2c\cite{caoyue1,Haijun2}. We note that Fig. 2b and 2c depict the overall dominant orbital textures that are deduced directly from the photoemission matrix element analysis of the light-polarization-dependent ARPES measurements\cite{caoyue1}.  A recent polarization-dependent ARPES measurements as well as density functional theory calculations on Bi$_2$Se$_3$ shows that the in-plane orbital textures may be layer-dependent\cite{Zhu}.

\noindent{\bf Proposed spin-orbital texture in Bi$_2$Se$_3$. }It was theoretically proposed that, the spin texture in the topological insulators is not only locked with the crystal momentum, but also coupled with the orbital texture\cite{Haijun2}. The spin texture of the topological insulators has been well studied theoretically\cite{Fu2,Basak} and experimentally\cite{Hsieh,Souma,Xu,Pan,Jozwiak,miyamoto}, leading to a usual picture as sketched in Fig. 2a that shows a left-handed spin texture for the upper Dirac cone and a right-handed spin texture for the lower one.  In fact, such a spin texture is not about the real spin but the total angular momentum \textbf{J} = \textbf{S} + \textbf{L}, which is the combination of the real spin \textbf{S} and the orbital angular momentum \textbf{L} due to the spin-orbital coupling\cite{Haijun2}.  Therefore, it is important to separate the pure spin contribution from the angular momentum of orbitals. Distinct spin textures are predicted for different orbital textures in the Bi$_2$Se$_3$ topological insulator (Fig. 2b and 2c)\cite{Haijun2}: (1). The out-of-plane {\it p}$_\text{z}$ orbital texture gives a left-handed helical spin texture for the upper Dirac cone and a right-handed one for the lower Dirac cone (Fig. 2b); (2). The in-plane tangential orbital texture gives a right-handed helical spin texture for the upper Dirac cone (Fig. 2c) and it is reversed for the lower Dirac cone (not shown in Fig. 2c); (3). The in-plane radial orbital texture gives a right-handed helical spin texture for the lower Dirac cone (Fig. 2c) that is reversed for the upper one (not shown in Fig. 2c). As shown above, the capability of selectively probing different components of the {\it p} orbitals by ARPES with polarization-variable light source makes it possible to detect the coupled spin-orbital texture when the spin detection capability is added. Given the dominant orbital textures determined in the {\it p} and {\it s} polarization geometries\cite{caoyue1,Haijun2}, it is possible to assign pure spin texture for the upper and lower Dirac cones as shown in Fig. 2b and 2c. For the {\it p} polarization geometry (Fig. 2b), the pure spin texture is opposite in its chirality for the upper and lower Dirac cones, similar to the total spin texture of the surface state (Fig. 2a). For the {\it s} polarization, however, the same spin texture is expected for both the upper and lower Dirac cones which is distinct from the total spin texture. These sharp predictions on the orbital-selective spin textures can be directly examined by SARPES with polarization-variable light source.

\noindent{\bf Direct observation of spin-orbital texture coupling in Bi$_2$Se$_3$.}  Figure 3 shows the measured results on the spin polarization of the Bi$_2$Se$_3$ Dirac surface state in the {\it s} polarization (Fig. 3a) and {\it p} polarization (Fig. 3b) geometries. Our newly-developed SARPES system based on VUV laser made it possible to carry out high-resolution spin-resolved photoemission measurements on both the upper and lower Dirac cones of the Bi$_2$Se$_3$ surface state that is critical for the present study (see Methods below).  In Fig. 3a with the {\it s} polarization geometry, the spin-resolved photoemission spectra (energy distribution curves, EDCs), labeled by S1 to S5 from bottom to top, were measured on five representative momentum points along the $\bar{\Gamma}\bar{K}$ cut as marked by the dashed lines in the band image measured from regular ARPES (Fig. 3c). Likewise, in Fig. 3b with the {\it p} polarization geometry, spin-resolved EDCs, labeled by P1 to P5, were measured on five momentum points along the $\bar{\Gamma}\bar{K}$ cut as marked by the dashed lines in the band image in Fig. 3d. At a given momentum point, the two EDCs represent the in-plane (sample surface plane) spin component along the vertical Z direction with up (red line plus circles) and down (blue line plus triangles) spin orientation (see Methods for details). There are three kinds of peaks observed in EDCs that correspond to the upper Dirac band (marked by black diamond), lower Dirac band (black asterisk), and bulk conduction band (black empty circle).  We note that the bulk conduction band also shows in-plane spin polarization along the vertical Z direction that exhibits interesting variations with momentum and the light polarization. In the present paper, we will mainly concentrate on studying the spin structure of the Dirac surface state and leave the discussion on bulk states elsewhere.

Let us now have a close examination on the spin polarization variation with momentum, energy and the light polarization. We start with the {\it s} polarization geometry measurements as shown in Fig. 3a and 3c.  In the spin-resolved EDCs for the momentum S5 corresponding to the right Fermi momentum k$_\text{FR}$ in Fig. 3c,  the peak located near the Fermi level (E$_\text{F}$) comes from the upper Dirac cone while the peak with 0.58 eV binding energy is from the lower Dirac cone.  At both of these two peak positions, the up-spin component dominates. This means that the dominant in-plane spin orientation points upward for both the points on the upper Dirac cone and the lower Dirac cone at this momentum k$_\text{FR}$, as marked in the right panel of Fig. 3c. When the momentum moves to S4 that is between k$_\text{FR}$ and the $\bar{\Gamma}$ point, there are three peaks observed in the spin-resolved EDCs: the peak close to the Fermi level is from the bulk conduction band, while the other two peaks around 0.28 eV and 0.44 eV come from the surface band above and below the Dirac cone, respectively. It is clear that these two peaks of the Dirac surface state show similar spin polarization behavior as observed at k$_\text{FR}$.  The momentum S3 is close to the $\bar{\Gamma}$ point and the up-spin EDC is nearly identical to the down-spin one, indicating that the in-plane spin polarization along the Z direction is very small.  After the momentum goes across the $\bar{\Gamma}$ point to the S2 and S1 points, the peak from the upper Dirac cone shows spin polarization again but with the opposite spin orientation when compared with the measurements for S5 and S4.  This means that the dominant spin direction for S1 and S2 is opposite to that of S4 and S5 for the upper Dirac cone.  The intensity of the lower Dirac cone at S1 and S2 points is relatively lower than that at S5 and S4 points, probably caused by the matrix element effects in the photoemission process.  It is clear that, for the S1 momentum point close to the left Fermi momentum k$_\text{FL}$ as marked in Fig. 3c, the down-spin component dominates for both the upper and lower Dirac cones, indicating that the dominant in-plane spin orientation points downward for both the upper and lower Dirac cones near k$_\text{FL}$ , as marked in the right side of Fig. 3c. The opposite spin-orientation at the left and right Fermi momenta is consistent with the unique spin-momentum locked spin texture in topological insulators.  Our SARPES results in the {\it s} polarization geometry (Fig. 3a and 3c) uncover, for the first time, an unexpected spin texture where the upper Dirac cone shares the same right-handed chirality with the lower one. Such a spin texture is unusual from the normal spin texture of topological insulators as usually considered (Fig. 2a).

The SARPES results of the Bi$_{2}$Se$_{3}$ surface state measured in the {\it p} polarization geometry are presented in Fig. 3b and 3d.  The experimental conditions for the {\it p} polarization geometry are similar to the {\it s} polarization geometry except for the light polarization change.  The momentum P5 is close to the right Fermi momentum k$_\text{FR}$ and thus at the similar momentum point with S5 in Fig. 3a and 3c. But there is a marked difference on the spin direction between these two light polarization geometries. While the dominant spin directions for the upper Dirac cone and the lower Dirac cone are the same in the {\it s} polarization geometry, they are opposite in the {\it p} polarization geometry. Specifically, for the upper Dirac cone, the spin direction measured by {\it p} and {\it s} polarizations is opposite while it is the same for the lower Dirac cone. A careful examination of the spin-resolved data at different momenta in Fig. 3b, as has been done for Fig. 3a for the {\it p} polarization geometry, suggests a spin texture picture shown in the right panel of Fig. 3d.   We note that, at P1  momentum point, the upper Dirac cone band is very weak and the peak near the Fermi level comes mainly from the bulk band, as shown in Fig. 3b and 3d. But at P2 and P3 momentum points, the bulk band and surface band are well separated and it shows clearly that the spin texture of the upper Dirac cone is opposite to that of the lower Dirac cone. For the {\it p} polarization geometry, the upper Dirac cone shows a left-handed chirality while the lower Dirac cone shows a right-handed chirality, which is similar to the total spin texture of the topological surface state in Bi$_{2}$Se$_{3}$.

We note that in the {\it s} polarization geometry, according to the photoemission matrix element analysis, the contribution from the radial orbital texture to the spectral intensity of the lower Dirac cone should be zero. The observed intensity of the lower Dirac cone is indeed strongly suppressed when comparing Fig. 3a measured in {\it s} polarization with Fig. 3b measured in {\it p} polarization geometry, consistent with the dominant radial orbital texture expected in the lower Dirac cone. However, there remains some weak signal present for the lower Dirac cone in the {\it s} polarization geometry (Fig. 3c); similar observation was also reported before\cite{caoyue1}.  The multi-component picture\cite{Haijun2,Zhu} seems to give a natural explanation of this residual spectral intensity when one considers there is a mixture of other components such as the tangential orbital texture, in addition to the dominant radial texture, that can give rise to this residual spectral weight. The layer-dependent orbital texture picture\cite{Zhu} has an additional advantage to explain the intensity asymmetry between the positive and negative momentum sides in {\it p} polarization (Fig. 3d). In addition, we can not fully rule out whether the residual spectral weight might be due to the finite momentum resolution or slight sample misalignment during the spin-resolved measurements.  The mixing of the bulk valence band with the lower Dirac band cannot be excluded either. The exact origin of the residual spectral weight needs further investigations.

In order to further confirm that the observed spin-texture switching is purely caused by the polarization change of the incident light, we carried out a particular SARPES experiment: we  measured on a Bi$_{2}$Se$_{3}$ sample by switching the light polarization only while keeping all the other experimental conditions exactly the same.  The spin-resolved EDCs, which were measured near the right Fermi momentum k$_\text{FR}$ marked by the the green line in the inset in Fig. 4a, show two peaks: the peak near the Fermi level corresponds to the upper Dirac cone while the peak at 0.60 eV binding energy is from the lower Dirac cone.  In the {\it s} polarization geometry (Fig. 4a), the upper Dirac cone is dominated by the up-spin component. When the light polarization is switched to {\it p} polarization geometry (Fig. 4b), the same upper Dirac cone becomes dominated by the down-spin component. On the other hand, the lower Dirac cone is dominated by the up-spin component in both the {\it s} and {\it p} polarization geometries.  We note that the spin polarization at the Dirac point clearly approaches to zero in both the {\it s} and {\it p} polarization geometry, reflecting the fact of Kramers degeneration at the Dirac point.  These results further corroborate the above measurements in Fig. 3. It also vividly demonstrates that the spin orientation of the upper Dirac cone can be manipulated up and down simply by switching the light polarization.\\

\noindent{\bf\large Discussion}

\noindent The change of the spin texture with the light polarization in the upper Dirac cone of the Bi$_2$Se$_3$ surface state was reported very recently by two independent spin-ARPES measurements\cite{caoyue2,Lanzara}. Quite different theoretical pictures were employed to interpret these similar experimental results. In one case, such a spin texture switching was attributed to the intrinsic property of the topological surface state in Bi$_2$Se$_3$ in terms of the spin-orbital texture coupling of the initial state\cite{caoyue2,Haijun2}. In the other case, it is interpreted as a result of the spin-dependent interaction of the helical surface electrons with the incident light, which originates from strong spin-orbit coupling and implies that the usually accepted spin-conservation scenario for spin-resolved photoemission may become invalid in these materials\cite{Lanzara}.
%In the other case, it is interpreted as a result of the spin-flipping transition due to the interaction of polarized light with photoelectrons during the photoemission process.
%This points to an extrinsic effect that the detected spin orientation does not represent the initial spin orientation of the Dirac surface state\cite{Lanzara}.
No distinction can be made between these two fundamentally different scenarios if only the upper Dirac cone is measured because they predict a similar spin texture for the upper Dirac cone. However, when the spin texture of both the upper Dirac cone and the lower Dirac cone is considered, it is possible to tell between them. In this case, according to the spin-orbital texture picture\cite{Haijun2}, the spin chirality of the lower Dirac cone can be either the same (in {\it s} polarization geometry, Fig. 2c) or opposite (in {\it p} polarization geometry, Fig. 2b) to that of the upper Dirac cone. On the other hand, according to the spin-dependent interaction interpretation\cite{Lanzara}, with the photoelectron spin texture predicted for the upper Dirac cone to be like in Fig. 4b%(changed from 4a_by shaolong)
(in {\it p}-polarization geometry) and 4c (in {\it s}-polarization geometry) of Ref. 17,  the expected spin orientation of the lower Dirac cone is always opposite to that of the upper Dirac cone, both in {\it p}- and {\it s}-polarization geometry.   It is therefore crucial to simultaneously detect the spin structure of both the upper Dirac cone and lower Dirac cone under different polarization geometries that has been a challenge for SARPES measurements\cite{miyamoto}. As shown in the above, our VUV laser-based state-of-the-art SARPES system (see Methods) made it possible to fulfill such measurements. Our observation of the same spin chirality for the upper and the lower Dirac cones in the {\it s} polarization geometry is not compatible with the spin-dependent interaction interpretation\cite{Lanzara} without considering the unique spin-orbital texture in the initial state.  Instead, the measured spin chirality and its relative orientation for the upper and lower dirac cones in both the {\it s} and {\it p} polarization geometries (right sides of Fig. 3c and 3d)  show a perfect agreement with that expected from the spin-orbital texture picture (Fig. 2b and 2c). In particular, the unusual observation of the same spin chirality for the upper and lower Dirac cones in the {\it s} polarization geometry can be naturally understood in this orbital-dependent spin texture picture as proposed by Zhang et al.\cite{Haijun2} and Zhu et al.\cite{Zhu}. However, we caution that we cannot completely exclude the contribution of the bulk effects that may hinder the clear observation of the spin polarization from the surface Dirac cone. In particular,  the contribution of the bulk states may get larger in the spin-resolved photoemission intensities in \emph{p}-polarization geometry due to the stronger hybridization between the lower surface Dirac cone and the bulk valence band\cite{Jozwiak}.

Our understanding of the observed light polarization-dependent spin texture in Bi$_2$Se$_3$ in terms of the spin-orbital texture\cite{Haijun2} seems to contradict from the
previous spin-dependent interaction interpretation\cite{Lanzara} where the spin orientation is varied by light polarization during the photoemission process due to spin-dependent interaction of the helical surface electrons with the incident light. In non-magnetic materials, the spin-flipping transition due to the direct coupling of the radiation field of the incident light is usually thought to be negligible when compared with the spin-conserving electric dipole transitions\cite{Feder}. In the presence of the spin-orbit coupling, the spin-conserving electric dipole transition can produce the spin-dependent transition matrix elements\cite{Park}.  However, we note that the predicted photoelectron spin texture for the {\it p}-polarization geometry shown in Fig. 4b of Ref. 17 is the same as the calculated initial spin texture coupled with $p_{x}$ orbital in Fig. 3a of Ref. 18. Similarly, the spin texture for the {\it s}-polarization geometry (see Fig. 4c of Ref. 17) is the same as the initial spin texture coupled with $p_{y}$ orbital (see Fig. 3b of Ref. 16). These similarities may not be surprising because the predicted photoelectron spin texture in Fig. 4 of Ref. 25 is based on the calculations that already included the matrix element effect associated with the photoemission process\cite{Park}. In this sense, even though no orbital is explicitly specified in Ref. 17, the spin-resolved ARPES still picks different partial orbital components of the total $p$ orbital when different linear light polarizations are used. Different spin textures of the photoemitted electrons are detected because they are coupled with the selected orbital textures of the initial states. This interpretation of the predicted spin texture in Ref. 17 seems more plausible than  %the spin-flipping picture.
the interpretation merely based the spin-dependent interaction of the helical surface electrons with the incident light without considering the spin-orbital textures coupling in Bi$_2$Se$_3$. It may also reconcile the calculated results in Ref. 16 and Ref. 18.  The unique spin texture of Bi$_2$Se$_3$ detected with different light polarizations can still be understood within the normal spin-conservation transitions and is a demonstration of the intrinsic spin-orbital texture coupling in topological insulators.

In conclusion, by taking high resolution SARPES measurements with variable light polarizations, which cover both the upper Dirac cone and the lower Dirac cone, we have revealed interesting spin textures in the Bi$_2$Se$_3$ topological insulator.  (1). In the {\it s} polarization geometry, both the upper and the lower Dirac cones have the same helical spin texture with the right-handed chirality. This is fundamentally different from the general spin texture of topological insulators; (2). In the {\it p} polarization geometry, the upper Dirac cone has a left-handed spin texture while the lower Dirac cone shows a right-handed spin texture;  (3). By varying between the {\it s} and {\it p} polarization geometry, the spin texture of the upper Dirac cone can be switched between the right-handed and the left-handed chirality while the lower Dirac cone keeps the same chirality. These observations constitute strong evidence of the spin-orbital texture in the Bi$_2$Se$_3$ topological insulator.  Our results also indicate that the light-polarization-dependent spin texture is the manifestation of the intrinsic spin-orbital texture in the initial state combined with the photoemission matrix element effects.  This restores the validity of the generally accepted spin-conservation picture that is a foundation of the spin-resolved photoemission technique.  Our new observation of the spin-orbital texture, in addition to its usual spin-momentum locking, provides an additional degree of freedom in controlling the spin structure in topological insulators. It also demonstrates that light manipulation of  spin texture is possible in topological insulators that is important for their future applications in spin-related technologies.\\

\noindent{\bf\large\textbf{Methods} }

\noindent{\bf Samples growth methods.} High quality single crystals of Bi$_{2}$Se$_{3}$ were grown by the self-flux method\cite{Chaoyu}. Bismuth and selenium powders were weighed according to the stoichiometric Bi$_{2}$Se$_{3}$ composition. After mixing thoroughly, the powder was loaded into an alumina crucible and sealed in a quartz tube. The above processes were all done in an argon-atmosphere glove box with O$_\text{2}$ $\leq 0.1\ ppm$ and H$_\text{2}$O $\leq 0.1\ ppm$. The quartz tubes were took out and sealed after being evacuated. The mixed materials were heated to 1000 $^\circ$C, held for 12 hours to obtain a high degree of mixing, and then slowly cooled down to 500 $^\circ$C over 100 hours before cooling to room temperature. Single crystals of nearly one centimeter in size were obtained by cleaving.\\

\noindent{\bf The  {\it s}  and {\it p} polarization geometries in photoemission measurements.} The experimental geometry for the sample, the electron energy analyzer and the light is shown in Fig. 5. The mirror plane is defined by the lens axis of the electron energy analyzer and the surface normal of the measured sample.  The direction of the \textbf{E} vector for the linearly polarized incident vacuum ultraviolet (VUV) laser can be continuously varied.  For the {\it s} polarization geometry (Fig. 5a), the  \textbf{E} vector (shown as the green arrow) is perpendicular to the mirror plane. For the {\it p} polarization geometry (Fig. 5b), the \textbf{E} vector lies within the mirror plane.\\

\noindent{\bf The spin- and angle-resolved photoemission methods.} The spin- and angle-resolved photoemission measurements were performed on our newly developed vacuum ultraviolet (VUV) laser-based state-of-the-art spin- and angle-resolved photoemission system (SARPES), which combines the Scienta R4000 analyzer with a Mott-type spin detector. The photon energy of the laser is 6.994 eV with a bandwidth of 0.26 meV. The best energy resolution for regular ARPES measurements is $\sim$1 meV.

The spin detection in our SARPES system is realized by using a Mott-type spin detector that consists of a heavy element target (thorium) with strong spin-orbital coupling, and four surrounding channeltrons to detect the scattered electrons (Fig. 6d). When the spin-polarized photoelectrons are accelerated by high voltage (25 KV) and hit the target, the intensity of the scattered electrons measured by the four channeltrons will be different due to the spin-orbital interaction between the incident electrons and nucleus in the target. Such a signal difference can be taken as a measure of the spin polarization of the incident electrons. In the Mott-type spin detector, each component of the spin polarization is detected by two channeltrons. For example, in our measurement geometry as shown in Fig. 6d, the vertical Z component of the spin polarization (P$_\text{Z}$) is determined by the left (L) and right (R) channeltrons while a pair of the up (U) and down (D) channeltrons measures the Y component (P$_\text{Y}$). It is well known that the Mott-type spin detector has extremely low efficiency leading to poor instrumental resolution and low detection efficiency for the normal SARPES measurements\cite{Dil1,Dil2}. The utilization of the VUV laser source has greatly improved the capability of our SARPES system. Due to the intrinsic narrow linewidth of the VUV laser ($\sim$0.26 meV) and its super-high photon flux\cite{Guodong}, we are able to obtain the best spin-resolved energy resolution of $\sim$2.5 meV (Fig. 6c). To our knowledge, this is the best energy resolution achieved so far in the spin-resolved photoemission measurements. The best angular resolution of our SARPES is $\sim$0.3 degree; the momentum resolution is further improved due to the utilization of a low photon energy of the VUV laser (h$\nu$=6.994 eV).  In addition, our SARPES system is able to do spin-resolved measurement and regular angle-resolved measurement simultaneously, thus achieving an accurate location of the measured momentum point.  To improve the data statistics and reduce the measurement time, we set the spin-resolved energy resolution at 25 meV for the SARPES measurements of Bi$_{2}$Se$_{3}$ reported in this work. The angular resolution used is 0.75/0.5 degree parallel/perpendicular to the momentum cut direction (corresponding to a momentum resolution of 0.023/0.015 $\AA^{-1}$ for the 6.994 eV photon energy).  The spin-resolving capability of our SARPES system is demonstrated by measuring a standard sample  Au(111) which serves as a good reference for the spin polarization and chirality measurements in the Bi$_2$Se$_3$ topological insulator. Due to the Rashba effect\cite{Rashba1,Rashba2}, the surface state of Au(111) splits into two branches of bands with  well-defined spin polarization (Fig. 6b)\cite{LaShell,Nicolay}. Two corresponding Fermi surface sheets are formed with well-defined spin texture (Fig. 6a): the inner cone has a left-handed chirality while the outer one has a right-handed chirality\cite{Dil1,Henk}. Figure 6e shows the band structure of the Au(111) surface state measured along a momentum cut shown as a pink thick line in Fig. 6a. The simultaneous measurement by the four channeltrons in the spin detector gives four spin-resolved photoemission spectra (EDCs) shown in Fig. 6f for the momentum point shown in Fig. 6e (dashed blue line).  Due to much improved energy and angular resolutions, two peaks are well resolved in our spin-resolved EDCs (Fig. 6f). There is little difference between the two EDCs from the up (U) and down (D) channels, indicating a negligible spin polarization along the Y direction that is an out-of-plane spin component. On the other hand, the two EDCs from the left (L) and right (R) channels show obvious difference in their intensity at two peak positions. Moreover, the relative intensity from these two channels is opposite for the two peaks: for the 0.16 eV binding energy peak associated with the outer cone, the L channel intensity (blue line) is higher than that of the R channel (red line);  for the 0.05 eV peak related with the inner cone, it is the opposite. This indicates that there exists spin polarization along the vertical Z axis that is in the plane of the Au(111) surface and  the spin polarization directions for these two peaks are opposite.  These observations are consistent with previous results on Au(111)\cite{Dil1,Henk} and demonstrate the spin-resolved capability of our SARPES system. In particular, it has established a good correspondence between the spin chirality and the intensity difference among the channeltrons, as marked in Fig. 6f.  For the inner cone in Fig. 6a with a left-handed chirality, it corresponds to the 0.05 eV peak where the right channel intensity is higher than that of the left channel, while for the outer cone with a right-handed chirality, it corresponds to the 0.16 eV peak where the intensity of the left-channel is higher than that of the right channel.

For the regular ARPES and spin-resolved ARPES measurements, the Fermi level is referenced by measuring on a clean polycrystalline gold that is electrically connected to the sample holder. The Bi$_{2}$Se$_{3}$ samples were all cleaved and measured at 30 K in vacuum with a base pressure better than 5$\times$10$^{-11}$ torr.\\

\noindent{\bf Analysis of the SARPES data.} Figure 7 shows the data analysis procedure for the spin-resolved EDCs in Figs. 3 and 4. In Fig. 7, both the original and processed SARPES data obtained in {\it s} (left panel) and {\it p} (right panel) polarization geometries are presented. The corresponding momentum point is near the right Fermi momentum k$_\text{F}$ as marked by the green line in the inset. Figure 7a and 7b show the original EDCs recorded by the left and right channels of the spin detector (see Fig. 6d) in {\it s } and {\it p } polarization geometry, respectively. According to the measurement geometry of the spin detector sketched in Fig. 6d, the in-plane (sample surface plane) spin polarization along Z axis, P$_\text{Z}$, can be determined by the intensity asymmetry between the left and right channels:

\begin{equation}\label{Eq.1}
  P_\text{Z} = \frac{1}{S_\text{eff}}\times\frac{I_\text{L}-I_\text{R}}{I_\text{L}+I_\text{R}},
\end{equation}
where I$_\text{L}$ and I$_\text{R}$ are intensities of the scattered photoelectrons collected by the left and right channels of the spin detector, respectively. S$_\text{eff}$ stands for the effective Sherman function\cite{Sherman,Dunning} and is known to be 0.17\cite{Berntsen,Burnett}. The obtained polarization curves, P$_\text{Z}$, show relatively high in-plane polarization of the upper and lower Dirac cones in Bi$_{2}$Se$_{3}$ (Fig. 7e and 7f). The values of P$_\text{Z}$ near the Fermi level are $\sim 54\%$ and $\sim 60\%$ measured in {\it s} and {\it p} polarization geometry, respectively. While the original EDCs recorded by the left and right channels in Fig. 7a and b clearly show nearly zero spin polarization around the Dirac point and above E$_\text{F}$, some data points of the calculated P$_\text{Z}$ curves deviate significantly from zero due to an overall low intensity and poor sign-to-noise ratio of the unpolarized background. We thus remove the data points above E$_\text{F}$ and  make a 5-points smooth to the P$_\text{Z}$ curves as presented in Fig. 7e and 7f for the {\it s} and {\it p} polarization geometry, respectively. With the polarization values P$_\text{Z}$  obtained, it is then straightforward to get the spin-resolved EDCs by the following equations:
\begin{equation}\label{Eq.2}
\begin{split}
 I_\text{Up}=\frac{1+P_\text{Z}}{2}\times (I_\text{L}+I_\text{R})  \\
 I_\text{Down}=\frac{1-P_\text{Z}}{2}\times (I_\text{L}+I_\text{R}),
\end{split}
\end{equation}
where I$_\text{Up}$ and I$_\text{Down}$ represent up- and down-spin components along Z axis. The obtained spin-resolved EDCs are shown in Fig. 7c and 7d for {\it s } and {\it p } polarization geometry, respectively which are plotted as Fig. 4 in the main text. The spin-resolved EDCs shown in Fig. 3a and 3b of the main text were obtained in a similar manner.

$^{*}$Correspondence and requests for materials should be addressed to X.J.Z. (XJZhou@aphy.iphy.ac.cn).\\

\noindent{\bf\large

\vspace{3mm}

\noindent {\bf\large Acknowledgement}\\
%\begin{acknowledgments}
The authors would like to thank H. J. Zhang, H. M. Weng, X. Dai, Z. Fang, D. S. Dessau, Z.-H. Zhu, A. Damascelli and A. Bansil for helpful discussions. This work is supported by the National Natural Science Foundation of China (91021006, 10974239 and 11174346) and the Ministry of Science and Technology of China (2011CB921703 and 2013CB921700).
%\end{acknowledgments}

%\noindent {\bf Supplementary Information}  is linked to the online version of the paper.

\vspace{3mm}

%%%\noindent {\bf Acknowledgements}

%%%%\vspace{3mm}

\noindent {\bf \large Author Contributions}\\
X.J.Z., Z.J.X. and S.L.H. proposed and designed the research.  Z.J.X., S.L.H., C.Y.C., Y.F., H.M.Y., A.J.L., L.Z., D.X.M., J.F.H., Y.Y.P., X.L., Y.L., G.D.L., X.L.D., L.Y., J.Z., S.J.Z., Z.M.W., S.S.Z., F.W., Q.J.P., X.Y.W., C.T.C., Z.Y.X. and X.J.Z contributed to the development and maintenance of Laser-ARPES system.  Z.J.X., S.L.H., C.Y.C., Y.F., H.M.Y. and A.J.L. carried out the experiment.  X.J.Z., S.L.H., Z.J.X. and C.Y.C. wrote the paper.

\vspace{3mm}

\noindent{\bf\large Additional information}\\
%\noindent{\bf Supplementary Information} accompanies this paper.

\noindent{\bf Competing financial interests:} The authors declare no competing financial interests.

\newpage

\newpage
\begin{figure}
\centering
\includegraphics[width=1.0\columnwidth,angle=0]{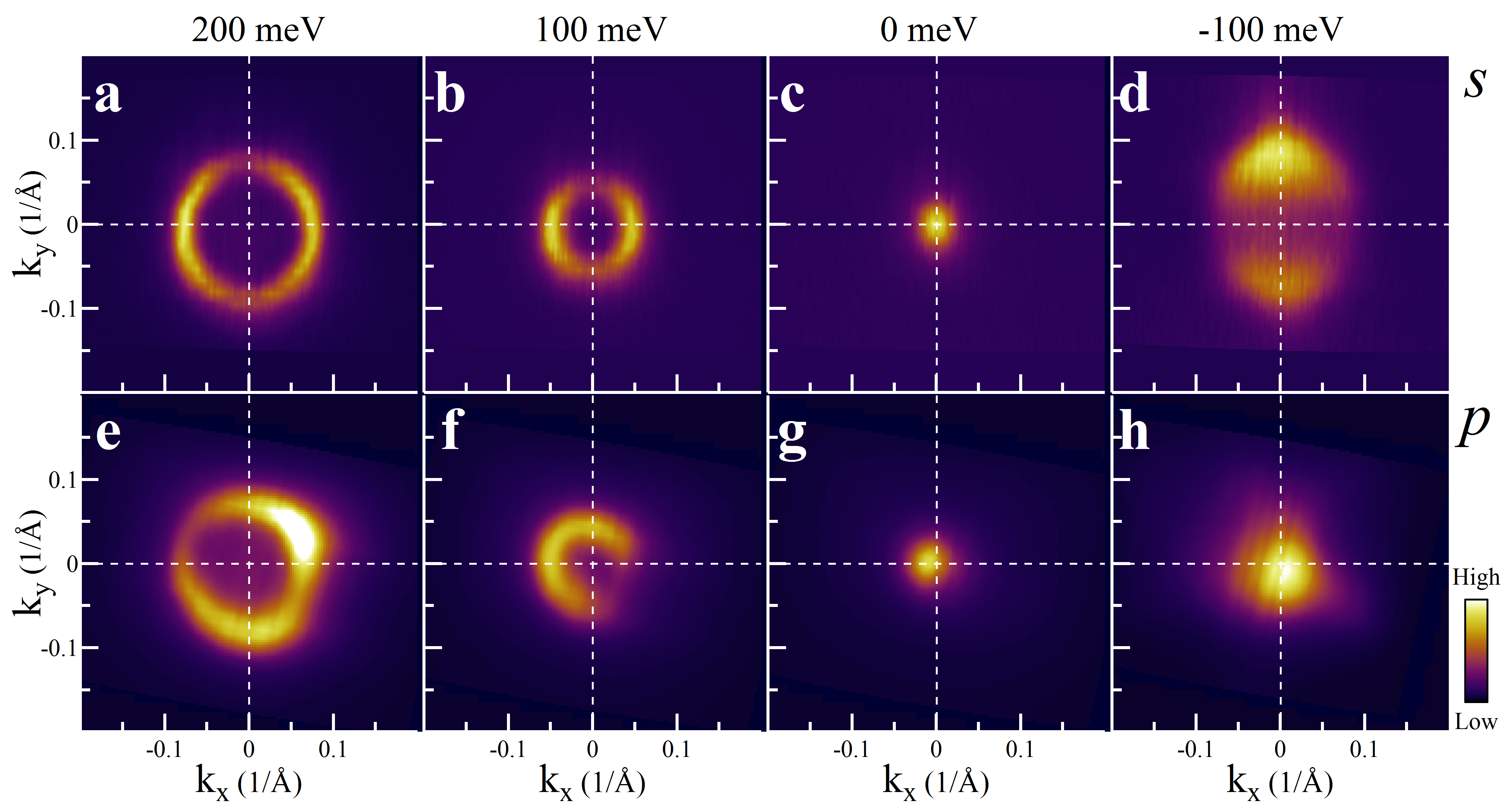}
\caption{\label{fig1}
{\bf ARPES intensity maps at different energies for the Bi$_2$Se$_3$ topological insulator.} The ARPES intensity maps measured in \emph{s}- and \emph{p}-polarization geometries are presented in the upper panels and lower panels, respectively. The energies labeled on the top of the figure are relative to the energy position of the Dirac point.  \textbf{a},\textbf{b}, \textbf{e}, and \textbf{f} are taken above the Dirac point, \textbf{c} and \textbf{g} are at the Dirac point, and \textbf{d} and \textbf{h} are below the Dirac point.}
\end{figure}

\begin{figure}
\centering
\includegraphics[width=1.0\columnwidth,angle=0]{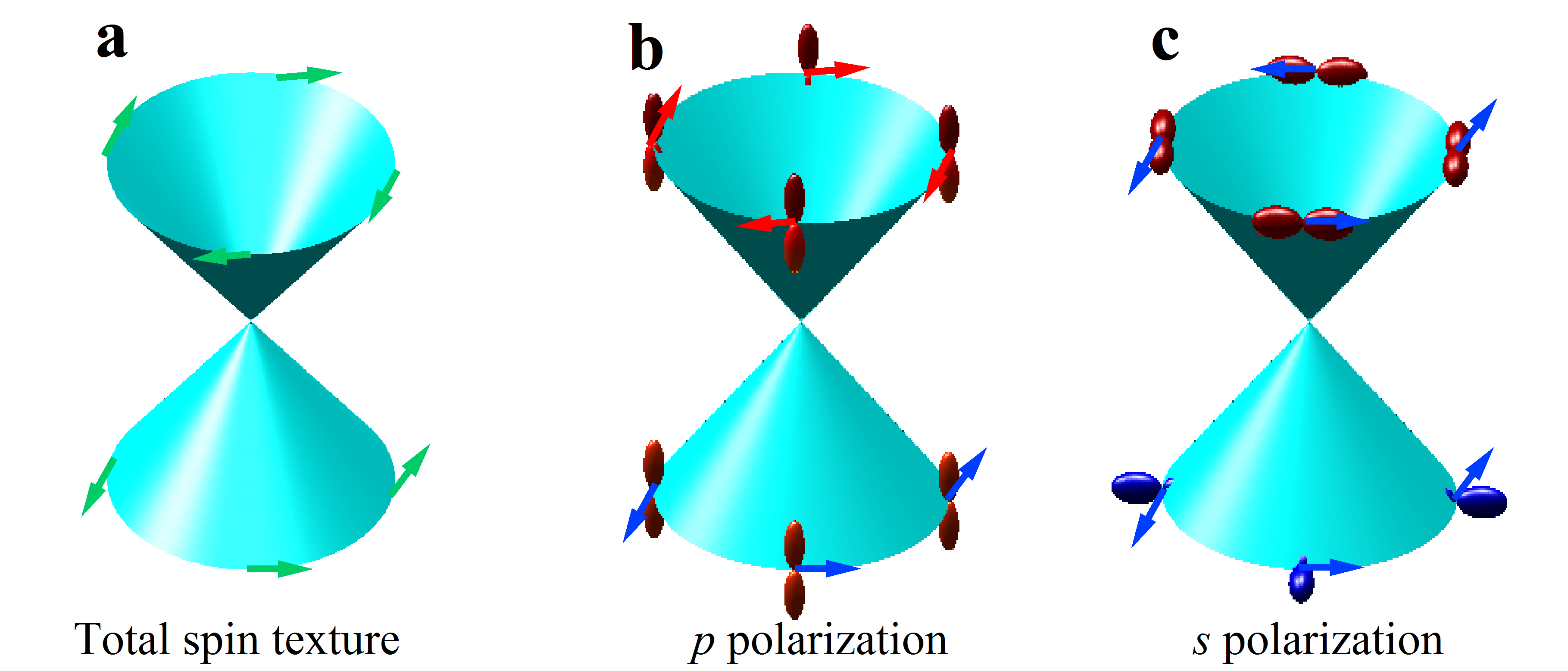}
\caption{\label{fig2}
{\bf Orbital textures and their coupled spin textures predicted for Bi$_2$Se$_3$.}
\textbf{a}. Schematic of the total spin texture for the surface state Dirac cone in topological insulators. The arrows indicate spin directions. \textbf{b}. The {\it p}$_\text{z}$ orbitals selectively probed in the {\it p} polarization geometry and their expected spin texture. In this geometry, from the analysis of the photoemission matrix element effect, the ARPES signal is dominated by the out-of-plane {\it p}$_\text{z}$ orbitals for both the upper Dirac cone and the lower Dirac cone\cite{caoyue1}. It is expected that the {\it p}$_\text{z}$ orbitals are associated with a left-handed helical spin texture for the upper Dirac cone and a right-handed spin texture for the lower Dirac cone\cite{Haijun2}.  \textbf{c}. Orbital textures probed in the {\it s}-polarization geometry and their related spin textures. In this geometry, the ARPES signal is dominated by the in-plane tangential orbital texture for the upper Dirac cone and the in-plane radial orbital texture for the lower Dirac cone\cite{caoyue1}. It is expected that the tangential orbital texture in the upper Dirac cone is associated with a right-handed helical spin texture and  the radial orbital texture in the lower Dirac cone is associated with a right-handed spin texture\cite{Haijun2}.
}
\end{figure}

\begin{figure}
\centering
\includegraphics[width=1.0\columnwidth,angle=0]{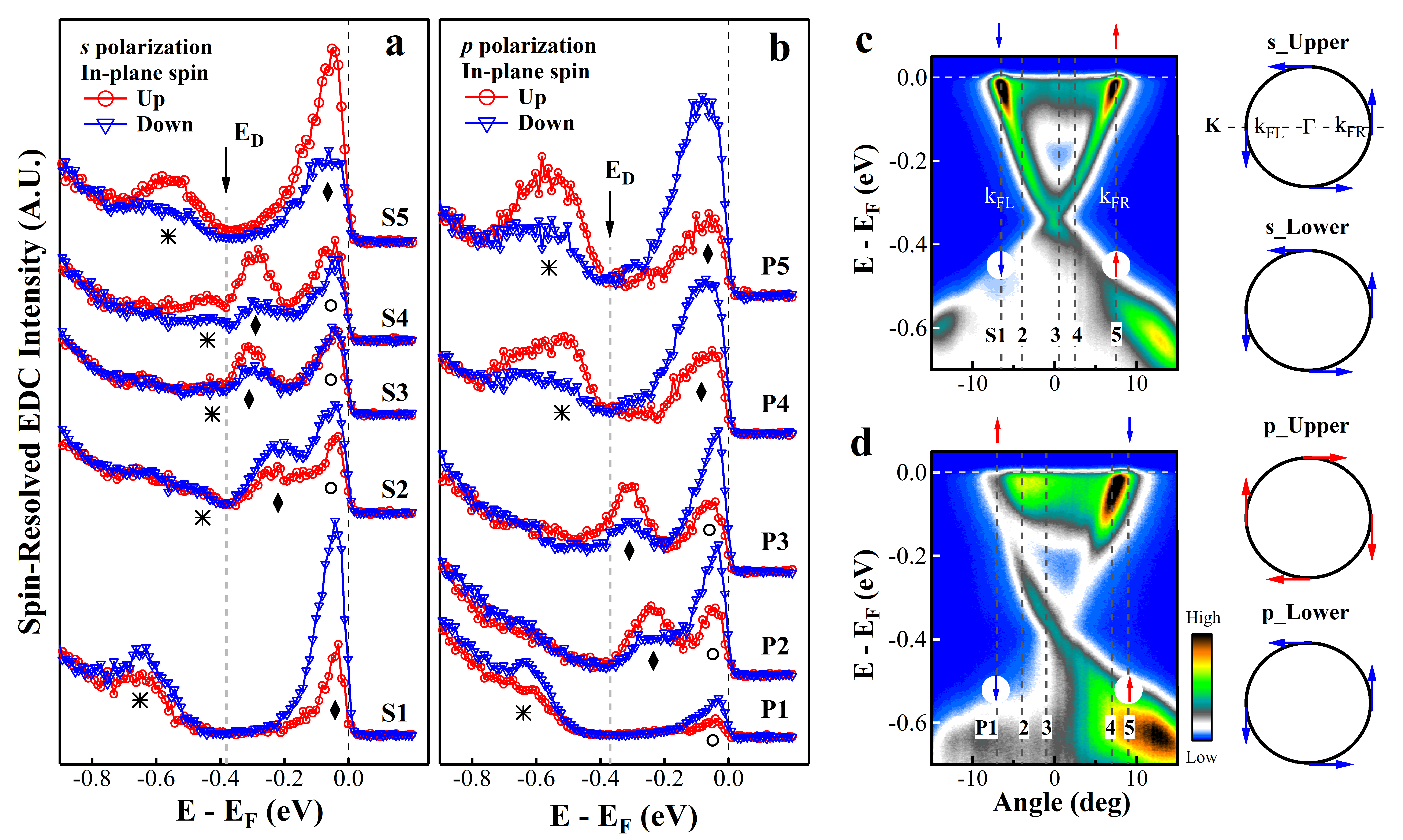}
\caption{\label{fig3} {\bf SARPES measurements of the Bi$_2$Se$_3$ topological insulator in different polarization geometries.} \textbf{a}. Spin-resolved EDCs at five representative momenta along the $\bar{\Gamma}\bar{K}$ momentum cut in the {\it s} polarization geometry. The two spin-resolved EDCs for a given momentum point represent two in-plane spin components along the vertical Z direction, one is up (red line plus circles) and the other is down (blue line plus triangles).  The corresponding momentum points are marked as the dashed lines in the band image measured from regular ARPES in the {\it s} polarization geometry (\textbf{c}). The EDC  peaks corresponding to the bulk band, the upper Dirac cone and the lower Dirac cone are marked by empty circle, solid diamond and asterisk, respectively.   \textbf{b}. Spin-resolved EDCs at five representative momenta along the $\bar{\Gamma}\bar{K}$ momentum cut in the {\it p} polarization geometry. The corresponding momentum points are marked as the dashed lines in the band image measured from regular ARPES in the {\it p} polarization geometry (\textbf{d}). On the right side of \textbf{c}, the measured spin texture in the {\it s} polarization geometry is sketched for both the upper Dirac cone and the lower Dirac cone. On the right side of \textbf{d}, the spin textures in the {\it p} polarization geometry is sketched.}
\end{figure}

\begin{figure}
\centering
\includegraphics[width=0.5\columnwidth,angle=0]{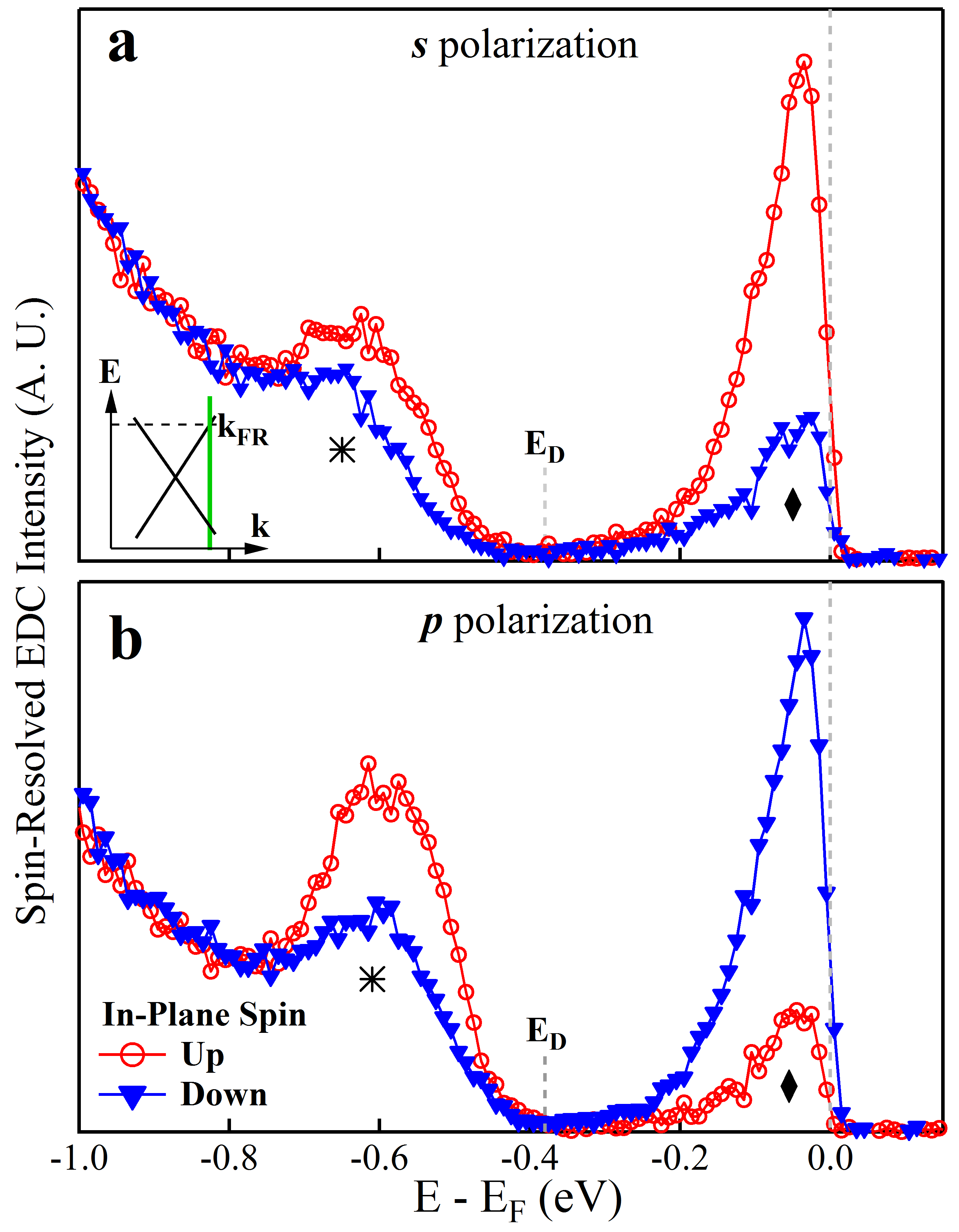}
\caption{\label{fig4}
{\bf Spin-resolved EDCs of the Dirac surface state in Bi$_2$Se$_3$ measured by only switching the light polarization.} When the light polarization state was switching from \emph{s} polarization to \emph{p} polarization, all the other experimental conditions are kept the same.  The corresponding momentum is near the right Fermi momentum k$_\text{FR}$ as marked by the green line in the inset.  The EDC  peaks corresponding to the upper Dirac cone and the lower Dirac cone are marked by solid diamond and asterisk, respectively. \textbf{a}. The spin-resolved EDCs measured in the {\it s} polarization geometry. \textbf{b}. The spin-resolved EDCs measured when the laser polarization is switched to the {\it p} polarization geometry.
}
\end{figure}

\begin{figure}
\centering
\includegraphics[width=1\textwidth]{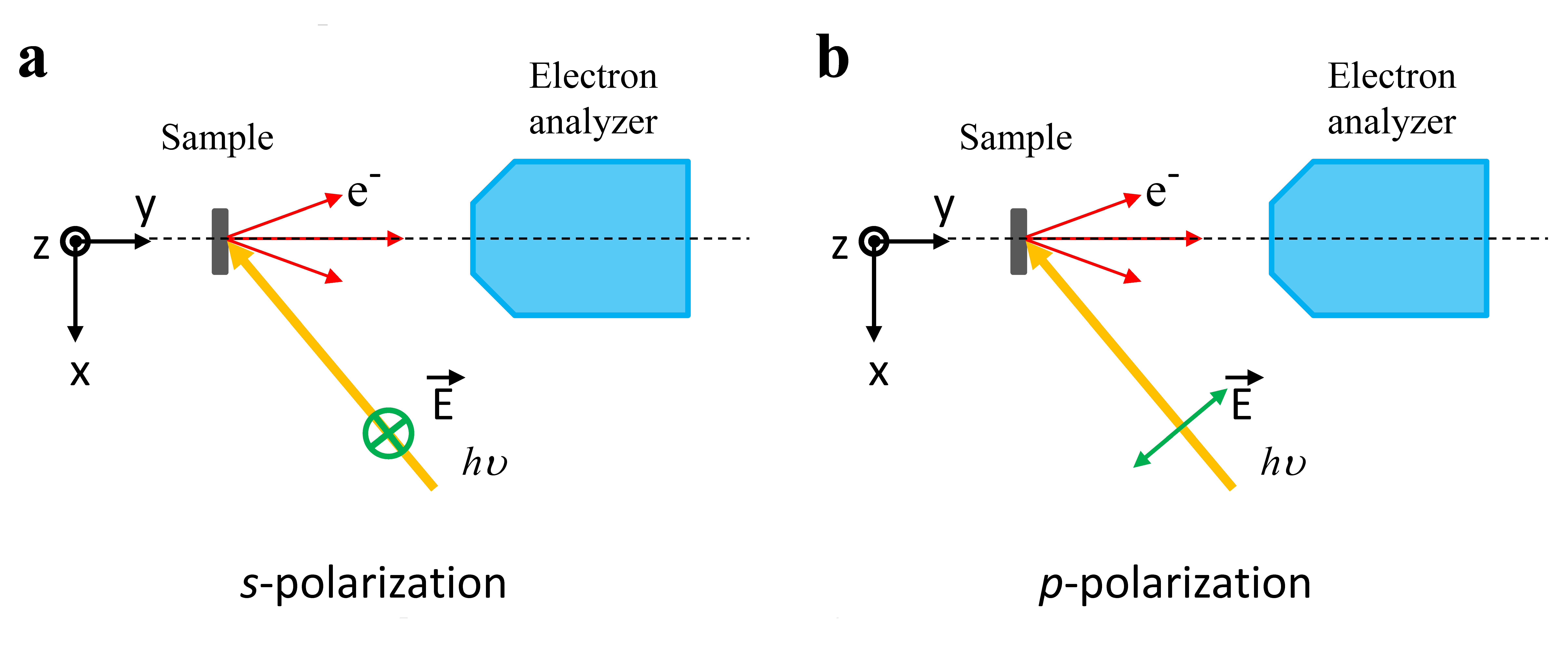}
\caption{\label{Fig5} {\bf Schematic experimental geometry of the {\it s} and {\it p} polarization in our SARPES measurements.} \textbf{a}. Schematic {\it s} polarization geometry where the laser electric field vector \textbf{E} (green arrow) is perpendicular to the mirror plane that is defined by the sample surface normal and the lens axis of the electron energy analyzer. \textbf{b}. Schematic {\it p} polarization geometry where the laser electric field vector \textbf{E} (green arrow) lies in the mirror plane.}
\end{figure}

\begin{figure}
\centering
\includegraphics[width=0.8\textwidth]{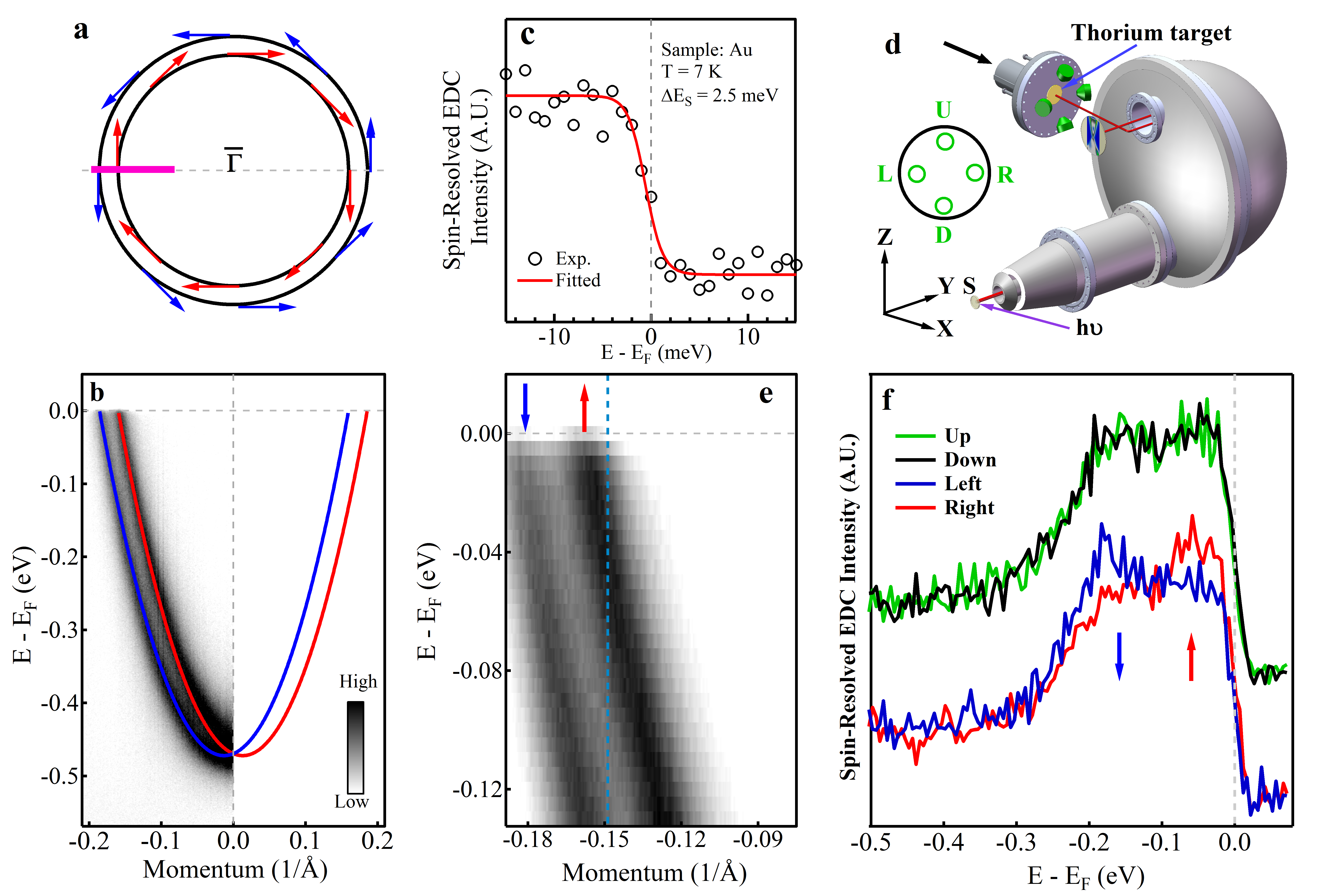}
\caption{\label{Fig6} {\bf The performance of the VUV laser-based spin- and angle-resolved photoemission system.}
\textbf{a}. Schematic Fermi surface and associated spin texture of the Au(111) surface state. The inner (outer) sheet shows a left-handed (right-handed) spin texture\cite{Dil1,Dil2}. \textbf{b}. Schematic band structure of the Au(111) surface state along the momentum cut shown in \textbf{a} as a dashed line. It is overlaid on the band structure measured by regular ARPES (left half of the momentum range).  \textbf{c}. Spin-resolved energy resolution test obtained by measuring the Fermi edge of a clean polycrystalline Au at 7 K. The measured data (open circles) are fitted by the Fermi distribution function (red solid line) and the overall fitted line width is 3.52 meV. By removing the thermal broadening, an instrumental spin-resolved energy resolution of 2.5 meV is obtained.  \textbf{d}. Schematic layout for our SARPES system which combines a Scienta R4000 electron energy analyzer with a Mott-type spin detector. There are four channeltrons in the spin detector to measure the scattered photoelectrons after they are accelerated and hit the thorium target. The four channeltrons are labeled in the inset viewed from the backside of the spin detector, as indicated by the black arrow.  In this geometry, the left and right channels measure the spin-polarization component along the vertical Z direction, while the up and down channels measure the spin-polarization component along the Y direction. \textbf{e} and \textbf{f} show, respectively, the band structure image and the four spin-resolved EDCs of the Au(111) surface state obtained simultaneously by the SARPES system.  The corresponding momentum cut for the image (\textbf{e}) is shown in (\textbf{a}) as a pink thick line.  The corresponding momentum point for the spin-resolved EDCs (\textbf{f}) is shown in (\textbf{e}) as a dashed blue line.
}
\end{figure}

\begin{figure}
\centering
\includegraphics[width=0.8\textwidth]{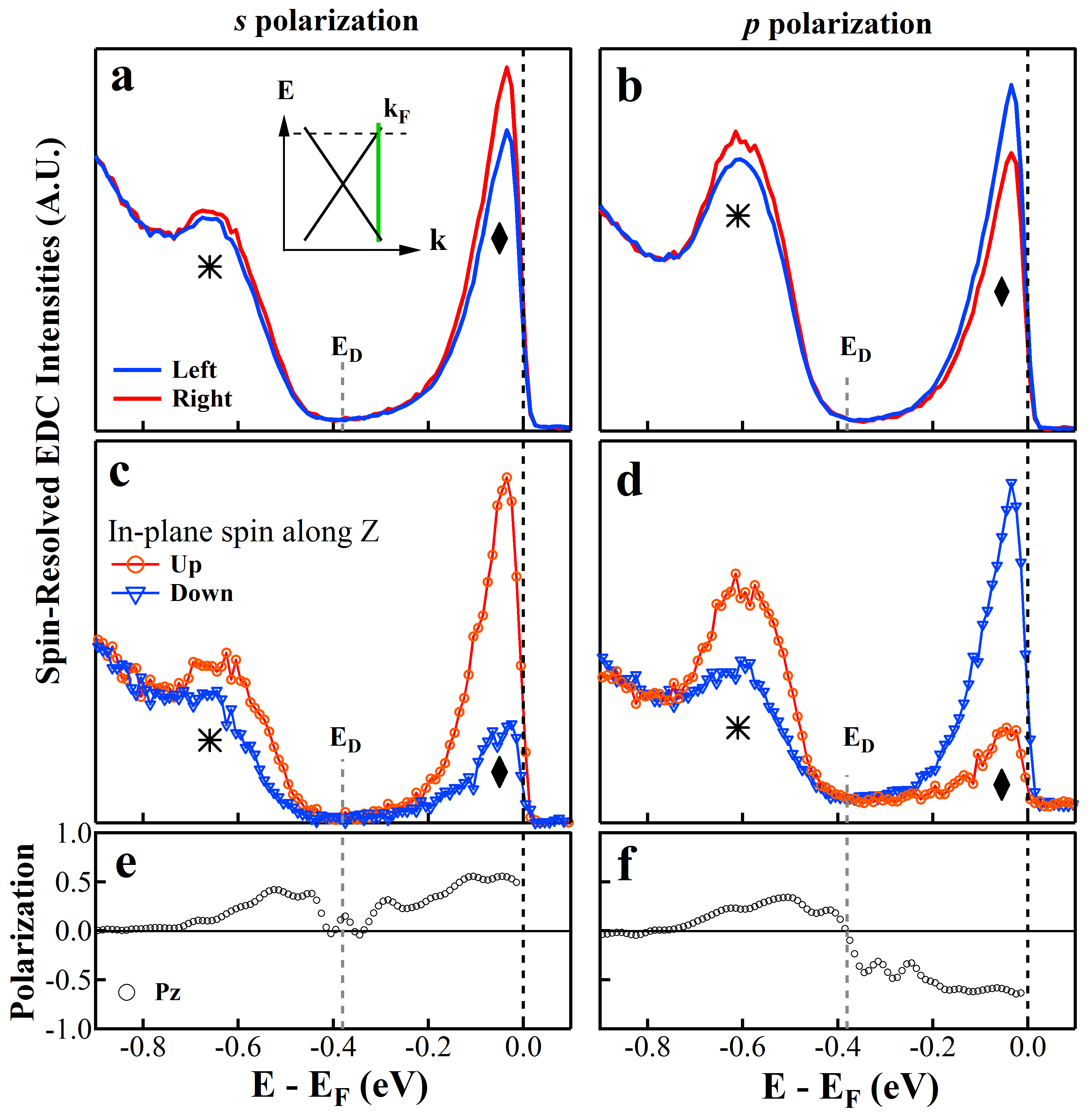}
\caption{\label{Fig7} {\bf Analysis procedure of the SARPES data of the  Bi$_2$Se$_3$ surface state measured in two different light polarization geometries.} The left/right panel presents the SARPES results obtained in {\it s}/{\it p} polarization geometry. The corresponding momentum point is near the right Fermi momentum k$_\text{F}$ as marked by the green line in the inset. The EDC  peaks corresponding to the upper Dirac cone and the lower Dirac cone are marked by solid diamond and asterisk, respectively. \textbf{a}/\textbf{b}. The original EDCs recorded by the left and right channels of the spin detector (see Fig. 6d) in {\it s}/{\it p} polarization geometry. \textbf{c} and \textbf{d}. The spin-resolved EDCs of the in-plane spin component along Z axis sketched in Fig. 6d. \textbf{e} and \textbf{f}. The in-plane electron polarization curves after a 5-points smooth. }

\end{figure}


\textbf{References}}
\begin{thebibliography}{99}

\bibitem{Fu1} Fu, L., Kane, C. L. and Mele, E. J. Topological insulators in three dimensions. {\it Phys. Rev. Lett. } {\bf 98}, 106803 (2007).
\bibitem{Qi1} Qi, X.-L. and Zhang, S.-C. The quantum spin Hall effect and topological insulators.{\it Phys. Today} {\bf 63}, 33 (2010).
\bibitem{Hasan1} Hasan, M. and Kane, C. Colloquium: Topological insulators. {\it Rev.  Mod. Phys.} {\bf 82}, 3045-3067 (2010).
\bibitem{Haijun1} Zhang, H. J. et al. Topological insulators in Bi$_2$Te$_3$, Bi$_2$Te$_3$ and Sb$_2$Te$_3$ with a single Dirac cone on the surface. {\it Nat. Phys.} {\bf 5}, 438-442 (2009).
\bibitem{Chaoyu} Chen, C. Y. et al. Robustness of topological order and formation of quantum well states in topological insulators exposed to ambient environment. {\it Proc. Natl Acad. Sci. USA} {\bf 109}, 3694-3698 (2012).
\bibitem{Zhangwei} Zhang, W. et al. First-principles studies of the three-dimensional strong topological insulators Bi$_2$Te$_3$, Bi$_2$Se$_3$ and Sb$_2$Te$_3$. {\it New. J. Phys.} {\bf 12}, 065013 (2010).
\bibitem{Fu2} Fu, L. Hexagonal warping effects in the surface states of the topological insulator Bi$_2$Te$_3$. {\it Phys. Rev. Lett.} {\bf 103}, 266801 (2009).
\bibitem{Basak} Basak, S. et al. Spin texture on the warped Dirac-cone surface states in topological insulators. {\it Phys. Rev. B} {\bf 84}, 121401 (2011).
\bibitem{Louie} Yazyev, O. V. et al. Spin polarization and transport of surface States in the topological insulators Bi$_{2}$Se$_{3}$ and Bi$_{2}$Te$_{3}$ from first principles. {\it Phys. Rev. Lett.} {\bf 105}, 266806 (2010)
\bibitem{Hsieh} Hsieh, D. et al. Observation of unconventional quantum spin textures in topological insulators. {\it Science}  {\bf 323}, 919-922 (2009).
\bibitem{Souma} Souma, S. et al. Direct measurement of the out-of-plane spin texture in the Dirac-cone surface state of a topological insulator. {\it Phys. Rev. Lett.} {\bf 106}, 216803 (2011).
\bibitem{Xu} Xu, S. Y. et al., Realization of an isolated Dirac node and strongly modulated Spin Texture in the topological insulator Bi$_2$Te$_3$. Preprint at http://arxiv.org/abs/1101.3985 (2011).
\bibitem{Pan}Pan, Z. H. et al. Electronic structure of the topological insulator Bi$_{2}$Se$_{3}$ using angle-resolved photoemission spectroscopy: evidence for a nearly full surface spin polarization. {\it Phys. Rev. Lett.} {\bf 106}, 257004 (2011).
\bibitem{Jozwiak} Jozwiak, C. et al. Widespread spin polarization effects in photoemission from topological insulators. {\it Phys. Rev. B} {\bf 84}, 165113 (2011).
\bibitem{miyamoto} Miyamoto, K. et al. Topological surface states with persistent high spin polarization across the Dirac point in Bi$_2$Te$_2$Se and Bi$_2$Se$_2$Te. {\it Phys. Rev. Lett.} {\bf 109}, 166802 (2012).
\bibitem{Park} Park, C. H. et al. Spin polarization of photoelectrons from topological insulators. {\it Phys. Rev. Lett.} {\bf 109}, 097601 (2012).
\bibitem{Lanzara} Jozwiak, C. et al. Photoelectron spin-flipping and texture manipulation in a topological insulator. {\it Nat. Phys.} {\bf 9}, 293-298 (2013).
\bibitem{Haijun2} Zhang, H. J.,  Liu, C. X. and Zhang, S.-C.  Spin-orbital texture in topological insulators.  {\it Phys. Rev. Lett.} {\bf 111}, 066801(2013).
\bibitem{Shen1} Damascelli, A., Hussain, Z. and Shen, Z.-X. Angle-resolved photoemission studies of the cuprate superconductors. {\it Rev.  Mod. Phys.} {\bf 75}, 473-541 (2003).
\bibitem{Hasan1st} Xia, Y. et al. Observation of a large-gap topological insulator class with a single Dirac cone on the surface. {\it Nat. Phys.} {\bf5}, 398-402 (2009).
\bibitem{YLChenScience} Chen, Y. L. et al. Experimental realization of a three-dimensional topological insulator, Bi$_2$Te$_3$. {\it Science} {\bf325}, 178-181 (2009).
\bibitem{caoyue1} Cao, Y. et al. In-plane orbital texture switch at the Dirac point in the topological insulator Bi$_2$Se$_3$. {\it Nat. Phys.} {\bf9}, 499-504 (2013).
\bibitem{Dil1} Dil, J. H. Spin and angle resolved photoemission on non-magnetic low-dimensional systems. {\it J. Phys.: Condensed matter} {\bf 21}, 403001 (2009).
\bibitem{Dil2} Heinzmann, U. and Dil, J. H. Spin-orbit-induced photoelectron spin polarization in angle-resolved photoemission from both atomic and condensed matter targets. {\it J. Phys.: Condensed matter.} {\bf 24}, 173001 (2012).
\bibitem{Zhu} Zhu, Z.-H. et al., Layer-by-layer entangled spin-orbital texture of the topological surface state in Bi$_2$Se$_3$.
   {\it Phys. Rev. Lett.} {\bf 110}, 216401 (2013).
\bibitem{caoyue2} Cao, Y. et al. Coupled spin-orbital texture in a protypical topological insulator. Preprint at http://arxiv.org/abs/1211.5998 (2012)
%\bibitem{Lanzara} Jozwiak, C. et al. Photoelectron spin-flipping and texture manipulation in a topological insulator. {\it Nat. Phys.} {\bf 9}, 293-298 (2013).
\bibitem{Feder} Feder, R. (ed.). Polarizated electrons in surface physics (World Scientific, 1985).
%\bibitem{Park} Park, C. H. et al. Spin polarization of photoelectrons from topological insulators. {\it Phys. Rev. Lett.} {\bf 109}, 097601 (2012).
\bibitem{Guodong} Liu, G. D. et al. Development of a vacuum ultraviolet laser-based angle-resolved photoemission system with a superhigh energy resolution better than 1 meV. {\it Rev. Sci. Instru.} {\bf 79}, 023105 (2008).

%\bibitem{Chaoyu} Chen, C. Y. et al. Robustness of topological order and formation of quantum well states in topological insulators exposed to ambient environment. {\it Proc. Natl Acad. Sci. USA} {\bf 109}, 3694-3698 (2012).
%\bibitem{Chaoyu} Chen, C. Y. et al. Robustness of topological order and formation of quantum well states in topological insulators exposed to ambient environment. Proceedings of the National Academy of Sciences {\bf 109}, 3694 (2012).
%\bibitem{Dil1} Dil, J. H. Spin and angle resolved photoemission on non-magnetic low-dimensional systems. {\it J. Phys.: Condensed matter} {\bf 21}, 403001 (2009).
%\bibitem{Dil2} Heinzmann, U. and Dil, J. H. Spin-orbit-induced photoelectron spin polarization in angle-resolved photoemission from both atomic and condensed matter targets. {\it J. Phys.: Condensed matter.} {\bf 24}, 173001 (2012).
%\bibitem{Guodong} Liu, G. D. et al. Development of a vacuum ultraviolet laser-based angle-resolved photoemission system with a superhigh energy resolution better than 1 meV. {\it Rev. Sci. Instru.} {\bf 79}, 023105 (2008).
\bibitem{Rashba1} Rashba, E. I. et al. Properties of semiconductors with an extremum loop. {\it Soviet physics, Solid State} {\bf 2}, 1109 (1960).
\bibitem{Rashba2} Bychkov, Y. A. and Rashba, E. I. Oscillatory effects and the magnetic susceptibility of carriers in inversion layers. {\it J. Phys. C: Solid State Physics} {\bf 17}, 6039 (1984).
%%\bibitem{Stormer} Stormer, H. L. et al. Energy Structure and Quantized Hall Effect of Two-Dimensional Holes. Physical Review Letters {\bf 51}, 126-129 (1983).
\bibitem{LaShell} LaShell, S., McDougall, B. A. and Jensen, E. Spin splitting of an Au(111) surface state band observed with angle resolved photoelectron spectroscopy. {\it Phys. Rev. Lett.} {\bf 77}, 3419 (1996).
\bibitem{Nicolay} Nicolay, G. et al. Spin$-$orbit splitting of the L-gap surface state on Au(111) and Ag(111). {\it Phys. Rev. B} {\bf 65}, 033407 (2001).
\bibitem{Henk} Henk, J. et al. Spin$-$orbit coupling in the L-gap surface states of Au(111): spin-resolved photoemission experiments and first-principles calculations. {\it J. Phys.: Condensed Matter} {\bf 16}, 7581 (2004).
%\bibitem{caoyue1} Cao, Y. et al.  In-plane orbital texture switch at the Dirac point in the topological insulator Bi$_2$Se$_3$. arXiv:1209.1016 (2012).
%\bibitem{Haijun2} Zhang, H. J., Liu, C. X. and Zhang, S.-C. Spin-orbital texture in topological insulators.  arXiv:1211.0762 (2012).
\bibitem{Sherman} Sherman, N. Coulomb scattering of relativistic electrons by point nuclei. {\it Phys. Rev.} {\bf 103},1601 (1965).
\bibitem{Dunning} Gay, T. J. Mott electron polarimetry. {\it Rev. Sci. Instrum.} {\bf 63}, 1635 (1992).
\bibitem{Berntsen} Berntsen M. H. A spin- and angle-reolving photoelectron spectrometer. {\it Rev. Sci. Instrum.} {\bf 81}, 035104 (2010)
\bibitem{Burnett} Burnett, G. C. et al. High-efficiency retarding-potential Mott polarization analyzer. {\it Rev. Sci. Instrum.} {\bf 65}, 1893 (1994).
\end{thebibliography}
\end{document}